\documentclass[aip, amsmath,amssymb, reprint]{revtex4-1}

\usepackage{graphicx}
\usepackage{dcolumn}
\usepackage{bm}
\usepackage[utf8]{inputenc}
\usepackage[T1]{fontenc}
\usepackage{mathptmx}
\usepackage{floatrow}

\begin{document}
\preprint{AIP/123-QED}

\title{Stabilization of exponential number of discrete remanent states with localized spin-orbit torques}

\author{Shubhankar Das}
\author{Ariel Zaig}
\author{Moty Schultz}
\author{Lior Klein}
\email{Lior.Klein@biu.ac.il}
\affiliation{Department of Physics, Nano-magnetism Research Center, Institute of Nanotechnology and Advanced Materials, Bar-Ilan University, Ramat-Gan 52900, Israel}

\date{\today}

\begin{abstract}
Using bilayer films of $\beta$-Ta/Ni$_{0.8}$Fe$_{0.2}$, we fabricate structures consisting of two, three and four crossing ellipses which exhibit shape-induced bi-axial, tri-axial and quadro-axial magnetic anisotropy in the crossing area, respectively. Structures consisting of N crossing ellipses can be stabilized in 2N remanent states by applying (and removing) an external magnetic field. However, we show that with field-free spin-orbit torques induced by flowing currents in individual ellipses, the number of remanent states grows to 2$^\text{N}$. Furthermore, when the current flows between the edges of different ellipses the number of remanent states jumps to 2$^\text{2N}$, including states which exhibit a $\pi$-N\'{e}el domain wall in the overlap area. The very large number of accessible remanent magnetic states that are exhibited by the relatively simple magnetic structures paves the way for intriguing spintronics applications including memory devices.
\end{abstract}

\maketitle
A promising way to address the rapidly rising demand for denser memory is storing more states in a single memory cell. This approach has led to the development of multi-level cell (MLC) FLASH memory with quad level cells\cite{Shibata}, MLC phase change memory \cite{Bedeschi,Athmanathan} and MLC resistive random access memory \cite{Xu_ReRAM,Liu_ReRAM}. Due to the unique advantages of magnetic random access memory (MRAM)\cite{Bhatti} which combines fast read and write operation, long retention and unlimited read and write endurance, it is important to develop also MLC magnetic memory for denser memory devices. Multi-level magnetic memory can be obtained by using 3D instead of planar structures \cite{J.Kim,Pirovano}, double/triple/quadrupole patterned lithography or multiple bit per cells\cite{Jameson}. Furthermore, if a sufficiently large number of states per cell is obtained, such memory cells would enable developing memristive logic computation\cite{Yang}, neuromorphic computing \cite{Querlioz,Lequeux,Grollier} and computing using high radix number systems \cite{Kim}.

Realizing a magnetic memory cell with multiple states is challenging, as magnetic shape anisotropy is commonly used to induce uniaxial anisotropy which supports only two remanent magnetic states. Higher order magnetic anisotropy cannot be realized in an entire magnetic structure solely with shape anisotropy; however, it has been shown that structures consisting of N crossing ellipses (CEs) give rise to N easy axes in the overlap area\cite{Telepinsky_JAP,Telepinsky_APL,Das_SR,Das_SR2}.  The direction of the easy axes is parallel to (in between)  the long axes of the ellipses when N is odd (even); however, away  from the overlap area there is only a single easy axis along the ellipse.  Such structures would normally support 2N remanent states which may be attractive for applications; however, clearly if the number of remanent states would grow exponentially with the number  of ellipses, it would be a dramatic improvement.

For practical applications of such structures an efficient and scalable method of switching between the different remanent states is required. Spin-orbit torques (SOTs) generated by flowing current in ferromagnetic/heavy metal (FM/HM) heterostructures owing to the spin Hall effect in HM  offer such a method  \cite{Chernyshov,Miron_Rashba,Miron_SHE,Liu_science,Liu_PRL,Pai,Fan,Yu,Hao,Hung,Aradhya,Fukami2,Fukami,Oh,Xu,Chen,Das_SR,Das_SR2,Baek}. Furthermore, SOTs have also been used very efficiently to systematically control domain walls motion\cite{Miron_DM1,Miron_DM2,Heinen,Haazen,Emori,Ryu} and for creation and manipulation of skyrmions including depinning and motion through nanostructures\cite{Jonietz,Jiang,Buttner,Legrand,Deger} for realizing next generation logic and memory devices.

In this study, we use $\beta$-Ta/NiFe bi-layers to fabricate structures consisting of two-crossing ellipses (2CE), three-crossing ellipses (3CE) and four-crossing ellipses (4CE) which exhibit in the overlap area of the ellipses bi-axial, tri-axial and quadro-axial magnetic anisotropy, respectively. We show that driving currents which induce SOTs along individual ellipses increases the number of remanent states that can be stabilized from 2N to 2$^\text{N}$. Furthermore, by flowing currents across the edges of different ellipses, the number of remanent states jumps to 2$^\text{2N}$ including states where the magnetization points at opposite directions in the edges of an individual ellipse, which gives rise to various magnetic configurations in the overlap area, including $\pi$-N\'{e}el-type domains. This means, that in case of a 4CE structure, instead of 8 states that can be stabilized by external magnetic fields, the manipulation of the same structure with SOTs that are affecting only part of the structure increases the number of remanent states to 256. The large number of states that can be achieved in relatively simple magnetic structures may pave the way to multi-level MRAM devices. Alternatively, it may be the basis for various type of memory devices, e.g. memristor with a history-dependent response which may make it suitable for a magnetic analog of neural computation.

\begin{figure}
\begin{center}
\includegraphics [trim=0cm 0cm 0cm 0cm, angle = 0, width=8.5cm,angle=0]{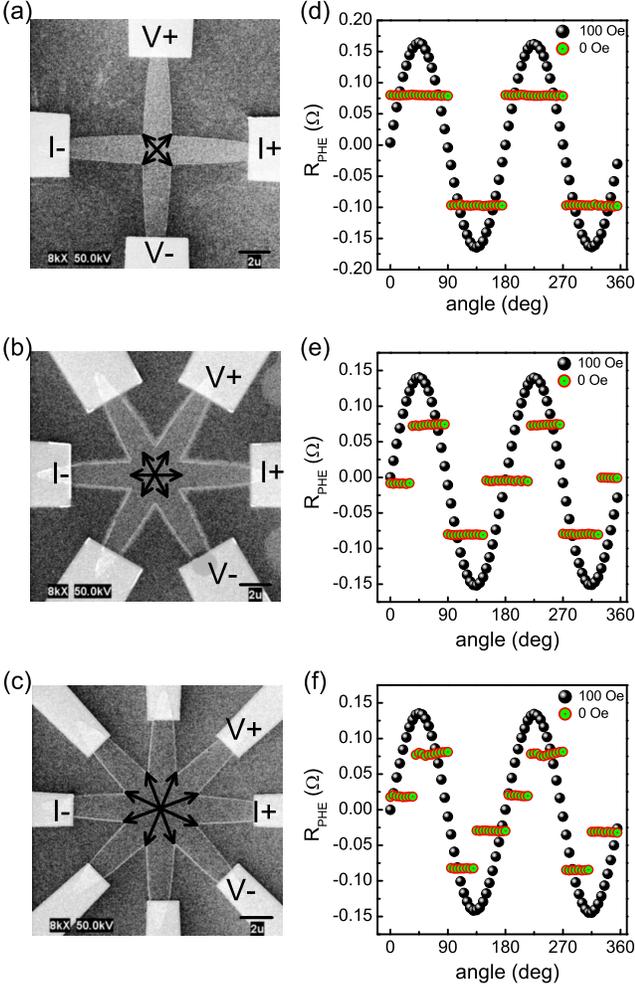}
\end{center}
\caption {\label{Devices} (a-c) Scanning electron microscopy (SEM) images of the 2CE, 3CE and 4CE, respectively. Current and voltage pads for planar Hall response measurements are indicated. The direction of all the easy axes are shown by the arrows. (d-f) R$_\text{PHE}$ as a function of field angle $\alpha$ for 2CE, 3CE and 4CE, respectively. The data are taken with a field of 100 Oe and after removing the field at each field angles.}
\end{figure}

Heterostructures of $\beta$-Ta(5 nm)/Ni$_{0.8}$Fe$_{0.2}$(2 nm)/Ti(3 nm) are deposited on thermally oxidized Si-wafer in an ion-beam sputtering system \cite{Cardoso}. Patterns of CEs, having principle axes of 2 and 16 $\mu$m, are fabricated using e-beam lithography followed by Ar-ion milling. Second stage e-beam lithography, Au sputtering and lift-off are performed for defining contact pads connected with all the edges of the ellipses. Scanning electron microscopy images of the devices are shown in Fig. \ref{Devices}(a), (b) and (c). Measurements are performed at room temperature using a home-made system consisting of Helmholtz coils and a sample rotator with angular precision of 0.03$^\circ$.

The magnetic configuration in the overlap area of the CEs is monitored by measuring the planar Hall resistance $R_{\text{PHE}} = \Delta V/I$ (see Fig. \ref{Devices}(a-c)) which for our polycrystalline samples is given by\cite{Ky}; R$_{\text{PHE}}$  = $\frac{1}{2}\Delta {\text{R}}\sin 2\theta$, where $\Delta {\text{R}}$ is the anisotropic magnetoresistance amplitude and $\theta$ is the angle between current and magnetization. Fig. \ref{Devices}(d-f) show R$_{\text{PHE}}$ as a function of the angle between the field and the current direction ($\alpha$) of 2CE, 3CE and 4CE, respectively, and for each $\alpha$, R$_{\text{PHE}}$ is measured with a saturating field of 100 Oe and after the field is switched off. The observed four, six and eight plateaus indicate  bi-axial, tri-axial and quadro-axial anisotropy of 2CE, 3CE and 4CE structures, respectively, with 2N remanent states. Please note that in the 4CE structure, $\Delta V$ is intentionally not measured across the perpendicular ellipse to avoid obtaining the same remanent value of R$_{\text{PHE}}$ expected for $\theta= \pi/8$ and  $\theta= 3\pi/8$ (see Fig. S1 of the supplementary materials).

\begin{figure*}
\floatbox[{\capbeside\thisfloatsetup{capbesideposition={right,top},capbesidewidth=4cm}}]{figure}[\FBwidth]
{\includegraphics [width=12.75cm]{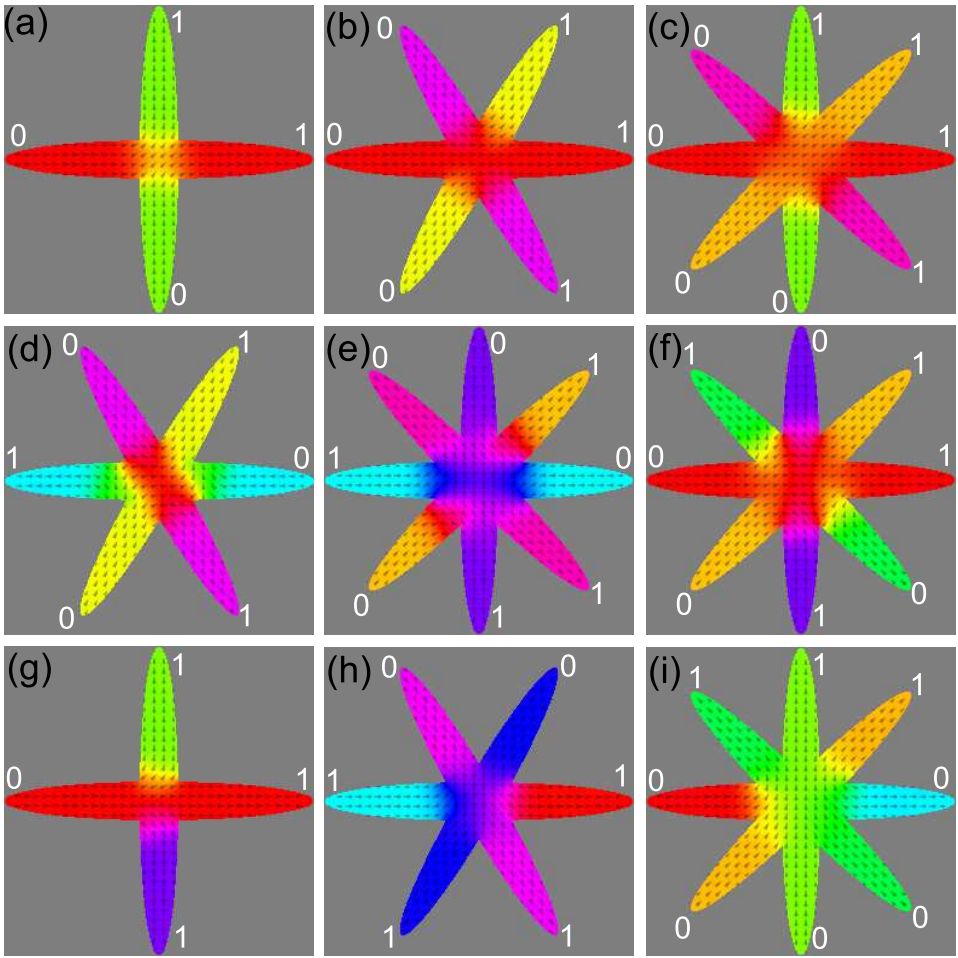}}
{\caption {\label{config_state} (a-c) Micromagnetic simulated OSs of 2CE, 3CE and 4CE, respectively, which can be stabilized by external magnetic field. (d-f) Simulated SSs generated by selectively manipulating the entire individual ellipses of 3CE and 4CE, respectively, which can not be stabilized by external magnetic field. (g-i) Simulated NSs generated by stabilizing the magnetization of both edges of individual ellipse in opposite direction for 2CE, 3CE and 4CE, respectively. The arrows indicate the direction of magnetization. Ones and zeroes at the edges of the ellipses indicate outward and inward magnetization, respectively.}}
\end{figure*}

As we show below, manipulating the CE structures with SOTs gives rise  to 2$^\text{2N}$ remanent states which can be divided into three types. Fig. \ref{config_state} presents the three remanent types obtained with numerical simulations ( Mumax3 \cite{Vansteenkiste}):  (i) ordinary remanent states (OS) which are the only accessible states when a uniform external magnetic field is applied (and removed) on the entire structure of CEs (Fig. \ref{config_state}(a-c)), (ii)  staggered remanent states (SS) accessible when the magnetization of each individual ellipse can be manipulated independently  (Fig. \ref{config_state}(d-f)), and (iii) $\pi$-N\'{e}el-like remanent states (NS) accessible only when the magnetization of each half of an individual ellipse can be manipulated independently (Fig. \ref{config_state}(g-i)).  Whereas the number of OS (2N) grows linearly with N, if SS and NS become accessible, the number of remanent states is given by  2$^\text{N}$ or  2$^\text{2N}$ respectively, namely, it grows exponentially with the number of ellipses.

A convenient representation of the different states is obtained  by assigning zero or one to the edges of each ellipse if the magnetization points inwards or outwards, respectively (see Fig. \ref{config_state}). Thus the state of a CE structure with N ellipses is described by a sequence of 2N numbers: a$_1$, a$_2$, ..., a$_\text{2N}$ which reflect the magnetization at the edges of the ellipses as we go around the structure. In the case of an OS the sequence should contain N sequential zeros or ones, which gives rise to 2N different sequences. If each ellipse can be reversed individually and SS are also available  there is no requirement for N sequential zeros or ones; however, since each ellipse has the same magnetization at its edges then if a$_i$ is zero, a$_{\text{(i+N)} mod \text{2N}}$ must be one and vice versa. Therefore, there are only N independent values which give rise to 2$^\text{N}$ states. The correlation between different items of the sequence is removed if the edges of each ellipse can be manipulated separately, which gives rise to 2$^\text{2N}$ different sequences.

\begin{figure}
\begin{center}
\includegraphics [trim=0cm 0cm 0cm 0cm, angle = 0, width=8.5cm,angle=0]{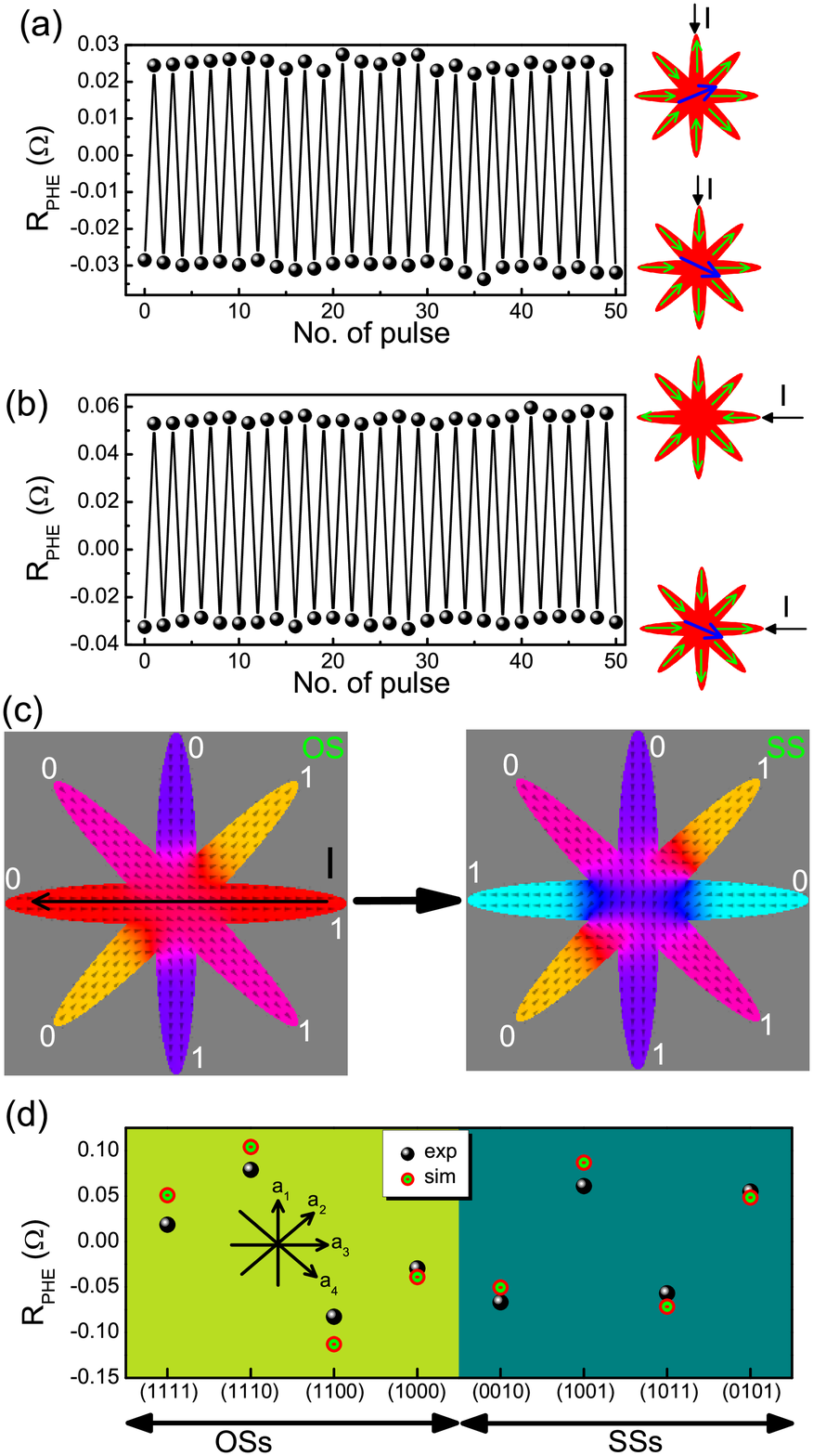}
\end{center}
\caption {\label{reversible switching_4el} (a) and (b) Field-free reversible switchings between two OSs and between an OS and a SS, respectively, by flowing current pulses of 2.5 mA through individual ellipse. The direction of magnetization of the two remanent states and the current direction are indicated by arrows in the schematics. (c) Micro-magnetic simulation of the switching between OS and SS by flowing a spin-polarized current in the horizontal ellipse. (d) The experimental R$_{\text{PHE}}$ of the OSs and the SSs along with numerical simulated values. The sequence of four number in the x-axis indicates the magnetization direction of the edges of the ellipses, where it starts from the top edge of vertical ellipse and rotates clockwise.}
\end{figure}

We start by showing how SS are obtained in a 4CE structure with SOTs.
Fig. \ref{reversible switching_4el}(a) and (b) show field-free SOT-induced switching experiments in such a structure by driving current pulses of 2.5 mA along one of the ellipses. Fig. \ref{reversible switching_4el}(a) demonstrates the switching between two plateau values of R$_{\text{PHE}}$ obtained in measurements as presented in  Fig. \ref{Devices}(f); namely, between OS, but in Fig. \ref{reversible switching_4el}(b) one of the values differs from all plateau values. The later state is identified as a SS based on both the numerical simulation of the switching and good agreement between the value of measured R$_{\text{PHE}}$ and the simulated R$_{\text{PHE}}$.

To simulate the switchings (Mumax3), we consider a 4CE device of dimension 2.048$\times$16.384 $\mu$m$^2$ and thickness 2 nm in an OS, where a spin-polarized current pulse of amplitude 8$\times$10$^{6}$ A/cm$^2$ and duration 2 ns is driven in the horizontal ellipse. The final state is achieved after relaxing the structure, which realizes the switching to a SS via reversing the magnetization of the same ellipse, shown in Fig. \ref{reversible switching_4el}(c). We use characteristic values of permalloy: Gilbert damping co-efficient of 0.01, saturation magnetization (M$_S$) = 8 $\times$ 10$^5$ A/m and exchange stiffness (A$_\text{ex}$) = 1.3 $\times$ 10$^{-11}$ J/m$^2$. Reversible switchings between other OSs and between OSs and SSs are shown in Fig. S2 and S3, respectively, of the supplementary materials.

To check whether the measured R$_{\text{PHE}}$ of the presumed SSs is consistent with the obtained magnetic configuration, numerical simulations are performed. We convert the magnetic configuration obtained by Mumax3 to conductivity tensor and using COMSOL Multiphysics R$_{\text{PHE}}$ is calculated. Experimental and simulated R$_{\text{PHE}}$ values of all the OSs and SSs are presented in Fig. \ref{reversible switching_4el}(d). The magnetic configurations are denoted by a sequence of four numbers since the remaining four numbers are the NOT values of shown four ones. The sequence starts from the top edge of vertical ellipse and rotates clockwise.

The stabilization of  SSs and switchings between  OS and SS in a 3CE structure are shown in Fig. S4 of supplementary materials.

\begin{figure}
\begin{center}
\includegraphics [trim=0cm 0cm 0cm 0cm, angle = 0, width=8.5cm,angle=0]{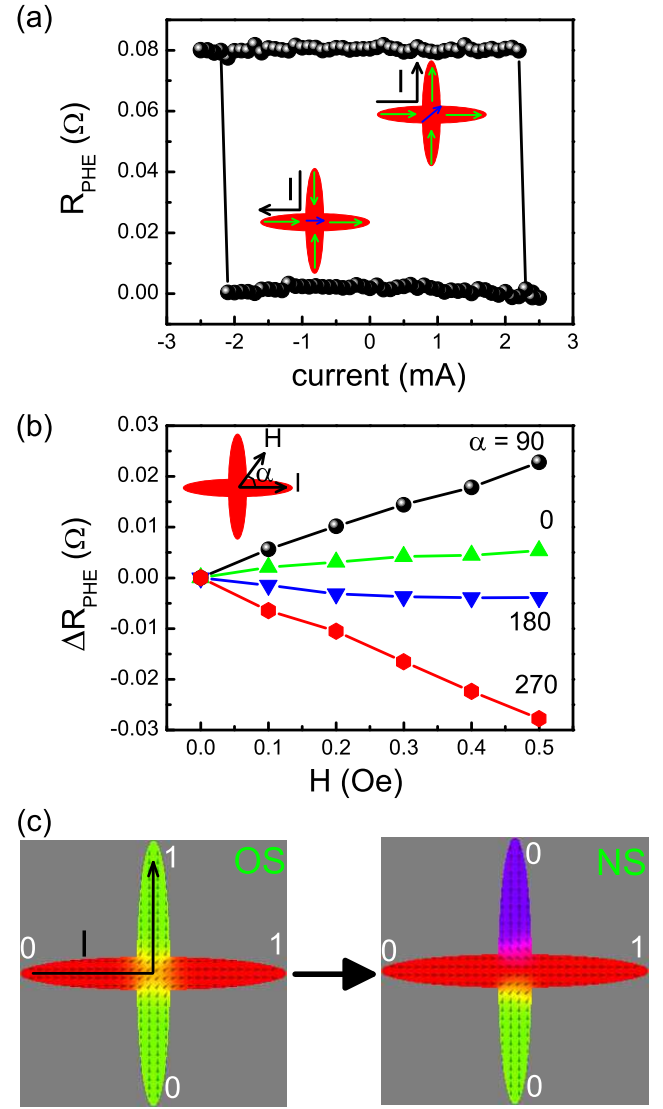}
\end{center}
\caption {\label{Neel_state} (a) Field-free reversible switching between an OS and a NS by flowing current between the edges of two different ellipses. The schematics show the current direction and the magnetization direction of both the edges and overlap area. (b) Change of R$_{\text{PHE}}$ as a function of field applied in various directions on a NS. (c) The switching in (a) is reproduced by micro-magnetic simulation by driving spin-polarized current as shown by the arrow.}
\end{figure}

We now turn to demonstrating the possibility to manipulate the two edges of each ellipse individually for achieving  2$^\text{2N}$ remanent states, and use for this purpose a 2CE structure. Starting from an OS of 2CE and flowing current between the edges of different ellipses (as shown in the schematics of Fig. \ref{Neel_state}(a)), a switching  occurs to a state with R$_{\text{PHE}}$ which differs from the values obtained for the OSs (see  Fig. \ref{Devices}(d)). To probe the magnetic configuration of the obtained state, we measure  R$_{\text{PHE}}$ vs H with field applied in various directions (see Fig. \ref{Neel_state}(b)). We observe a significant change in R$_{\text{PHE}}$ when the field is along $\alpha$ = 90 or 270 deg, whereas the change is minimal for $\alpha$ = 0 and 180 deg. This indicates that the magnetization in the overlap area is mostly along 0 deg. This is only possible if one of the two edges of the vertical ellipse reverses its magnetization, which in turn stabilizes a $\pi$-N\'{e}el domain wall in the overlap area. Micro-magnetic simulations using spin-polarized current pulse of amplitude 4$\times$10$^6$ A/cm$^2$ and duration 2 ns in a 2CE structure of dimension 2.048$\times$16.384 $\mu$m$^2$ and thickness 2 nm  supports the scenario of generating a  $\pi$-N\'{e}el-type domain wall (Fig. \ref{Neel_state}(c)). Flowing current selectively between the edges of different ellipses, all other NSs are stabilized and identified (see Fig. S5, S6 and S7 of supplementary materials). Namely, a 2CE structure can be effectively manipulated with SOTs between 2$^\text{2N}$ = 16 states.

Here we focus on demonstrating the ability of stabilizing exponential number of remanent states in relatively simple structures including intriguing N\'{e}el-like states. Clearly, the potential of these states for basic research and applications is yet to be explored. Thus for instance, using such structures for memory applications  would probably require a more efficient probing method than the planar Hall resistance which yields small difference in the resistance values obtained for different magnetic configurations and in addition it is symmetric under magnetization reversal. A possible route we are currently exploring is incorporating CEs structures as bottom ferromagnetic layers in magnetic tunnel junctions on top of a Ta layer, so that the CEs structures can be manipulated with SOTs. This would be useful not only for lifting the degeneracies under magnetization reversal but also for obtaining larger signals as the variation in MTJ resistance is orders of magnitude larger. We note that other methods to fabricate multi-level memory are based on a single axis of easy magnetization and the multiple states are achieved either by stacking MTJs vertically\cite{Zhang} or by moving a domain wall in one of the ferromagnetic layers between pinning sites\cite{Lequeux}. We believe that the method presented here is more appropriate for achieving a large number of discrete magnetic states.

Another issue relevant for memory applications is density. The structures presented here are on micron scale; however, based on previous works such as the demonstrated ability to stabilize N\'{e}el domain walls in NiFe nanowires of width 200-300 nm with symmetric anti-notches\cite{Oshea,Lage}, we expect that our structures may be downscaled by at least an order of magnitude. Furthermore, based on preliminary results N = 4 is not the upper limit for achieving well behaved structures. We also note that  multi-state memory devices particularly if the number of states can be as large as in our case may also be useful for memories that can be useful for neuromorphic computation.

In summary, structures consisting of two, three and four crossing ellipses support only 2N number of remanent states when stabilized by an external magnetic field. Here we show that using SOTs, it is possible to manipulate the magnetization of each ellipse individually and even each of its edges which enables stabilization of 2$^\text{2N}$ remanent states, including states which exhibit a $\pi$-N\'{e}el domain wall in the overlap area. The large number of states per structure and the ability of switching between them by SOTs in absence of any external magnetic field make such structures promising for various spintronics applications including memory devices.

See supplementary material for detailed experimental demonstration of switchings between different kinds of remanent states.

L. K. acknowledges support by the Israel Science Foundation founded by the Israel Academy of Sciences and Humanities (533/15).

The data that support the findings of this study are available from the corresponding author upon reasonable request.

\end{document}